\documentclass[preprint,showpacs,preprintnumbers,amsmath,amssymb]{revtex4}

\usepackage{graphicx}
\usepackage{dcolumn}
\usepackage{bm}

\begin{document}

\noindent

\preprint{}

\title{Quantization from an exponential distribution of infinitesimal action}

\author{Agung Budiyono}
\email{agungby@yahoo.com}

\affiliation{Jalan Emas 772 Growong Lor RT 04 RW 02 Juwana, Pati, 59185 Jawa Tengah, Indonesia}

\date{\today}

\begin{abstract} 

A statistical model of quantization based on an exponential distribution of infinitesimal action is proposed. Trajectory which does not extremize the action along an infinitesimal short segment of path is allowed to occur with a very small probability following an exponential law. Planck constant is argued to give the average deviation from the infinitesimal stationary action. 
 
\end{abstract}

\maketitle
 
\section{Motivation}  
   
As early as 1926 Madelung has shown that if one writes the complex-valued wave function into polar form, then the Schr\"odinger equation can be decomposed into a modified Hamilton-Jacobi equation and a continuity equation for a probability flow \cite{Madelung paper}. This observation has inspired many efforts to develop statistical models that lead to the derivation of the Schr\"odinger equation \cite{Fenyes,Weizel,Kershaw,Nelson,dela Pena,Santamato,Frieden,Garbaczewski,Reginatto,Kaniadakis,Hall,Markopoulou,Parwani,Smolin,Santos,Rusov,Gogberashvili}. It also provides the basis for the development of interpretation of quantum theory in term of traditional classical statistical mechanics. These include the hydrodynamics interpretation \cite{Madelung paper,Takabayasi,Janossy,Griffin,Wilhelm,Wong,Gosh,Sonego} and Bohmian mechanics \cite{Bohm,Holland,Bohm-Vigier,Valentini,Duerr,Wiseman}. Let us note however that the idea that the velocity of the particle is determined by the phase of a quantum wave in Bohmian mechanics is firstly developed by de Broglie as early as 1923. See for example the interesting historical description in Ref. \cite{Bacciagaluppi}.      
  
Further, despite the prominent role of Planck constant in the canonical quantization through replacement of c-number (classical number) with q-number (quantum number/Hermitian operators), its physical and dynamical origin, after more than one century since its first identification which signifies the birth of quantum theory, is still not clear. This question is of course beyond the standard formalism of quantum mechanics and can only be discussed if quantization is shown to arise effectively from a deeper theory. Hence, the elucidation of the physical origin of canonical quantization and thus of Planck constant might be indispensable for the searches of violations of quantum mechanics \cite{Bialynicki,Shimony,Ellis,Smolin VOQM,Weinberg,Bertolami,Gamboa,Hooft,Markopoulou VOQM,Adler,Perez,Valentini VOQM}, and may also lead to useful physical insights in the attempts to search for new physics within Planck scale, in which gravitational effect is no more ignorable.    

On the other hand, in the previous work \cite{AgungSMQ1}, we have developed a quantization method for systems of spin-less particles based on replacement of c-number by c-number parameterized by an unbiased random variable $\lambda$ to modify the classical Hamilton-Jacobi and continuity equations. The resulting modified Hamilton-Jacobi and continuity equations can then be rewritten into the Schr\"odinger equation when the distribution of $\lambda$ takes the form 
\begin{equation}
P(\lambda)=\frac{1}{2}\delta(\lambda-\hbar)+\frac{1}{2}\delta(\lambda+\hbar).  
\label{God's coin}
\end{equation}
Hence, $\lambda$ is an unbiased binary random variable which can take values only $\pm\hbar$. We then read-off a unique quantum Hamiltonian from the Schr\"odinger equation. Unlike canonical quantization, the method does not suffer from the problem of operator ordering ambiguity and the quantization processes (replacement) has direct and explicit interpretation as a specific modification of classical dynamics of ensemble of trajectories (in configuration space) parameterized by an unbiased random variable $\lambda$. 
 
In Ref. \cite{AgungSMQ0}, we have attempted to provide an argumentation that the rules of replacement postulated in Ref. \cite{AgungSMQ1} can be derived from a Hamilton-Jacobi theory with a specific random constraint uniquely determined by the classical Lagrangian. In the present paper, we shall give an alternative derivation of the rules of replacement postulated in Ref. \cite{AgungSMQ1} from a distinct statistical model based on {\it the assumption of exponential distribution of infinitesimal action}. Trajectory which does not extremize the action along an infinitesimal short segment of path is allowed to occur with a very small probability following an exponential law characterized by Planck constant. The Schr\"odinger equation and canonical commutation relation are shown as the implication of the model rather than postulated. Moreover, the configuration of the system evolves continuously in time and its effective velocity is related to the wave function in formally the same fashion as in Bohmian mechanics \cite{Bohm,Holland,Bohm-Vigier,Valentini,Duerr,Wiseman}. Yet, unlike the latter, the wave function is not physically real. 

\section{Exponential distribution of infinitesimal action}

Let us assume that in microscopic time scale the dynamics of the system is effectively stochastic. Further let us assume that the velocity of the system depends on a random variable $\lambda$, $\dot{q}\doteq dq/dt=\dot{q}(q,\lambda;t)$. Let us then consider an action along a segment of path connecting two infinitesimally close (configuration) spacetime points as:
\begin{equation}
\mathcal{I}(\lambda)=\int_{q(t)}^{q(t+dt)}L(q,\dot{q}(q,\lambda;t))dt,  
\label{action}
\end{equation}
where $L$ is the classical Lagrangian which now depends on $\lambda$ through $\dot{q}(q,\lambda;t)$. Let us assume that $\lambda$ is fixed along the short segment of path. This allows us to apply the principle of stationary action in the usual fashion which prescribes that the admissible classical trajectory connecting the two points is the one which extremizes the action. Solving $\delta\mathcal{I}(\lambda)=0$ with variation that vanishes at the end points, one obtains the Euler-Lagrange equation $(d/dt)(\partial L/\partial\dot{q})-\partial L/\partial q=0$. Let us assume for simplicity that the Lagrangian is not singular, $\mbox{det}\{\partial^2 L/\partial\dot{q}_i\partial\dot{q}_j\}\neq 0$. Then the Euler-Lagrange equation can be directly rewritten into the Hamilton equation 
\begin{equation}
\dot{q}\big(q,\underline{p}(q,\lambda;t);t\big)=\frac{\partial\underline{H}}{\partial\underline{p}},\hspace{2mm}\dot{\underline{p}}(q,\lambda;t)=-\frac{\partial\underline{H}}{\partial q}, 
\label{Hamilton equation}
\end{equation}
where $\underline{p}(q,\dot{q}(q,\lambda;t);t)\doteq\partial L/\partial\dot{q}$ is the conjugate canonical momentum and $\underline{H}(q,\underline{p})\doteq\underline{p}\dot{q}-L$ is the classical Hamiltonian which now depends on $\lambda$ through $\underline{p}(q,\lambda;t)$.   

Let us proceed to go to the Hamilton-Jacobi formalism through a canonical transformation so that the new classical Hamiltonian is vanishing. One then gets \cite{Goldstein book}
\begin{eqnarray}
-\underline{H}(q,\underline{p})=\partial_t\underline{S},
\label{Hamilton-Jacobi equation} 
\end{eqnarray}
where $\underline{S}(q,\lambda;t)$ is the generating function of the canonical transformation satisfying 
\begin{equation}
\underline{p}=\partial_q\underline{S}. 
\label{Hamilton-Jacobi condition}
\end{equation}
Inserting Eq. (\ref{Hamilton-Jacobi condition}) into the left hand side of Eq. (\ref{Hamilton-Jacobi equation}) one obtains the Hamilton-Jacobi equation $\partial_t\underline{S}+\underline{H}(q,\partial_q\underline{S})=0$. 

Equations (\ref{Hamilton-Jacobi equation}) and (\ref{Hamilton-Jacobi condition}) are equivalent to $d\underline{S}=\partial_t\underline{S}dt+\partial_q\underline{S}\cdot dq=-\underline{H}dt+\underline{p}\cdot dq=Ldt$ so that $d\underline{S}$ is the infinitesimal action along the classical (stationary) path. Now let us define a new function $S(q,\lambda;t)$ so that the difference along an infinitesimal stationary path is given by $dS=d\underline{S}=Ldt=-\underline{H}dt+\underline{p}\cdot dq$. Using $S$, the classical dynamics of the system along an infinitesimal segment of path with a fixed value of $\lambda$ can then be written in probabilistic form as: the probability that an infinitesimal segment of path occurs with $dS$ is given by 
\begin{equation}
P_S(d S|d\underline{S})\sim\delta(dS-d\underline{S})=\delta\big(dS-(\underline{p}\cdot dq-\underline{H}dt)\big), 
\label{principle of stationary action}
\end{equation}  
where $\delta$ now refers to the delta function. 
 
Written in the above form, one can then statistically modify the classical dynamics by allowing $dS$ to fluctuate around $d\underline{S}=\underline{p}\cdot dq-\underline{H}dt$. Let us then assume that there is an infinitesimal (microscopic) time scale $dt=\tau_Q$ so that the system can take a non-classical (non-stationary) segment of path with a value of $dS$, whose probability of occurrence is determined by its deviation from $d\underline{S}$ according to an exponential law as follows:
\begin{equation}
P_S(d S|d\underline{S})\sim e^{-\frac{2}{\lambda}(d S-d\underline{S})}e^{-\theta(S)d t}, 
\label{postulate of exponential distribution}
\end{equation}
where $\theta(S)$ is a function of $S$ evaluated at the initial point of the segment of the trajectory, whose form, as will be discussed below, is determined uniquely by the classical Hamiltonian. Evidently, $\lambda$ can not be vanishing and has to have the dimension of action. We assume that $dS\ge d\underline{S}$ for $\lambda>0$ and conversely, when $\lambda<0$, one assumes $dS\le d\underline{S}$. Classical dynamics is thus regained in the regime when $|d\underline{S}/\lambda|\gg 1$ or formally when $|\lambda|\ll 1$. In this case, $P_S(d S|d\underline{S})$ approaches a delta function centered at $dS=d\underline{S}$ of Eq. (\ref{principle of stationary action}). $|\lambda|$ thus gives the average deviation of $dS$ from $d\underline{S}$. 

Let us discuss how the assumption put in Eq. (\ref{postulate of exponential distribution}) modifies pair of Eqs. (\ref{Hamilton-Jacobi equation}) and (\ref{Hamilton-Jacobi condition}). Let us first denote the joint-probability density that the configuration of the system is $q$ with the value of the random variable $\lambda$ at time $t$  by $\Omega(q,\lambda;t)$. The marginal probability densities are then given by 
\begin{equation}
\rho(q;t)\doteq\int d\lambda\Omega\hspace{2mm}\&\hspace{2mm}P(\lambda)\doteq\int dq\Omega, 
\label{marginal probabilities general}
\end{equation} 
where we have assumed that the probability density of $\lambda$ is stationary (independent of time). 

Given a fixed value of $\lambda$, let us consider two infinitesimally close spacetime points $(q;t)$ and $(q+d q;t+d t)$. Let us assume that for this value of $\lambda$, the two points are connected to each other by  a segment of trajectory with $dS=\partial_tSdt+\partial_qS\cdot dq$. Then, for a fixed value of $\lambda$, according to the conventional probability theory, the probability density that the system initially at $(q;t)$ traces the segment of trajectory and end up at $(q+dq;t+dt)$, denoted below as $\Omega\big((q+dq,\lambda;t+dt)\big\|(q,\lambda;t)\big)$, is equal to the probability that the configuration of the system is $q$ at time $t$, $\Omega(q,\lambda;t)$, multiplied by the probability of occurrence of the segment of trajectory which is given by Eq. (\ref{postulate of exponential distribution}). One thus has 
\begin{eqnarray}
\Omega\Big((q+d q,\lambda;t+d t)\big\|(q,\lambda;t)\Big)\sim \Omega(q,\lambda;t)\nonumber\\
\times e^{-\frac{2}{\lambda}(d S-d\underline{S})}e^{-\theta(S)d t}.  
\label{probability density}
\end{eqnarray}  

Expanding the exponential on the right hand side up to the first order one gets $\Omega\big((q+d q,\lambda;t+d t)\|(q,\lambda;t)\big)\approx \big[1-(2/\lambda)(d S-d\underline{S})-\theta(S)d t\big]\Omega(q,\lambda;t)$. This can be rewritten as 
\begin{eqnarray}
d\Omega(q,\lambda;t)=-\Big[\frac{2}{\lambda}(d S-d\underline{S})+\theta(S)d t\Big]\Omega(q,\lambda;t), 
\label{fundamental equation 0}
\end{eqnarray} 
where $d\Omega(q,\lambda;t)\doteq\Omega\big((q+d q,\lambda;t+d t)\|(q,\lambda;t)\big)-\Omega(q,\lambda;t)$ is the change of the probability density $\Omega(q,\lambda;t)$ due to the transport along the segment of trajectory. 

Taking the limit $S\rightarrow\underline{S}$, Eq. (\ref{fundamental equation 0}) reduces into $d\Omega(q,\lambda;t)=-\Omega(q,\lambda;t)\theta(\underline{S})d t$. Dividing both sides by $d t$ and taking the limit $d t\rightarrow 0$, one obtains
\begin{eqnarray}
\frac{d\Omega}{dt}+\theta(\underline{S})\Omega=\partial_t\Omega+\dot{q}\cdot\partial_q\Omega+\theta(\underline{S})\Omega=0. 
\label{continuity equation 0}
\end{eqnarray}
To have a smooth correspondence with classical dynamics of ensemble, the above equation which describes the ensemble of classical trajectories, has to be equivalent to the classical continuity equation given by 
\begin{equation}
\partial_t\Omega+\partial_q\cdot(\dot{q}(\underline{S})\Omega)=\partial_t\Omega+\dot{q}(\underline{S})\cdot\partial_q\Omega+\partial_q\cdot\dot{q}(\underline{S})\Omega=0, 
\label{classical continuity equation}
\end{equation}
where the functional form of $\dot{q}$ with respect to $\underline{S}$ is given by substituting Eq. (\ref{Hamilton-Jacobi condition}) into the left equation of (\ref{Hamilton equation}). Comparing Eqs. (\ref{continuity equation 0}) and (\ref{classical continuity equation}), $\theta(\underline{S})$ thus has to be identified as the divergence of the classical velocity field
\begin{equation}
\theta(\underline{S})=\partial_q\cdot\dot{q}(\underline{S})=\partial_q\cdot\Big(\frac{\partial\underline{H}}{\partial\underline{p}}\Big|_{\underline{p}=\partial_q\underline{S}}\Big). 
\label{classical velocity field} 
\end{equation}  
Accordingly, it is sufficient (while not necessary) to assume that the functional form of $\theta(S)$ in Eq. (\ref{postulate of exponential distribution}) is given by replacing $\underline{S}$ in Eq. (\ref{classical velocity field}) with $S$
\begin{equation}
\theta(S)=\partial_q\cdot\Big(\frac{\partial\underline{H}}{\partial\underline{p}}\Big|_{\underline{p}=\partial_qS}\Big). 
\label{effective velocity divergence}
\end{equation} 

Let us go back to Eq. (\ref{fundamental equation 0}). Writing $d\Omega$ and $dS$  as $dF=\partial_tF dt+\partial_qF\cdot dq$, recalling $d\underline{S}=\partial_t\underline{S}dt+\partial_q\underline{S}\cdot dq=-\underline{H}dt+\underline{p}\cdot dq$, and comparing term by term one finally obtains  
\begin{eqnarray}
\underline{p}(q,\dot{q})=\partial_qS(q,\lambda;t)+\frac{\lambda}{2}\frac{\partial_q\Omega}{\Omega},\hspace{8mm}\nonumber\\
-\underline{H}(q,\underline{p}(q,\dot{q}))=\partial_tS(q,\lambda;t)+\frac{\lambda}{2}\frac{\partial_t\Omega}{\Omega}+\frac{\lambda}{2}\theta(S). 
\label{fundamental equation rederived}
\end{eqnarray} 
This is just the rules postulated in Ref. \cite{AgungSMQ1} where formal ``replacement'' there is shown here as physical ``substitution'' (see Eq. (5) of Ref. \cite{AgungSMQ1}). It is evident that as expected, in the formal limit $\lambda\rightarrow 0$, one regains (\ref{Hamilton-Jacobi condition}) and (\ref{Hamilton-Jacobi equation}) respectively. The above pair of equations can thus be regarded as the generalization of Hamilton-Jacobi theory. 

To see how Eq. (\ref{fundamental equation rederived}) modifies the classical dynamics of ensemble of trajectories, one thus needs to combine it with Eq. (\ref{classical continuity equation}). Inserting the upper equation of (\ref{fundamental equation rederived}) into the left hand side of the lower equation, one obtains a modified Hamilton-Jacobi equation. On the other hand, inserting the upper equation of (\ref{fundamental equation rederived}) into Eq. (\ref{classical continuity equation}) one gets a modified continuity equation. This is already shown in Ref. \cite{AgungSMQ1} to reproduce the results of canonical quantization for a wide class of classical systems of point-like particles with no spin if $\Omega(q,\lambda;t)=\rho(q,|\lambda|;t)P(\lambda)$ with $P(\lambda)$ is assumed to have the form given by Eq. (\ref{God's coin}). 

Next, inserting the first equation of (\ref{fundamental equation rederived}) into the left equation of (\ref{Hamilton equation}), one gets 
\begin{equation}
\dot{q}=\dot{q}(q,\underline{p};t)=\dot{q}(q,\partial_qS(q,\lambda;t)+\frac{\lambda}{2}\frac{\partial_q\Omega}{\Omega};t). 
\end{equation}
For example, if the classical Hamiltonian takes the form $H(q,\underline{p})=\underline{p}^2/(2m)+V(q)$ one has $\dot{q}(\lambda)=\underline{p}/m=\partial_qS/m+(\lambda/2m\Omega)\partial_q\Omega$. We shall show in the next section that for the general case when the classical Hamiltonian is at most quadratic in momentum, $\theta(S)$ given in Eq. (\ref{effective velocity divergence}) can be interpreted as the divergence of an effective velocity defined by averaging $\dot{q}(\lambda)$ over pair of opposite signs of $\lambda$. The exponential term $e^{-\theta(S)d t}$ in Eq. (\ref{postulate of exponential distribution}) thus describes whether the segment of trajectory effectively repels ($\theta>0$) or attracts ($\theta<0$) the nearby trajectories. Intuitively, the probability of occurrence of a segment of trajectory that effectively repels (attracts) the nearby trajectories is lower (higher). The probability of occurrence of a segment of classical trajectory with $dS=d\underline{S}$ is thus given by $P_S(d\underline{S}|d\underline{S})\sim e^{-\theta(\underline{S})d t}$. 
   
\section{A worked example: Hamiltonian with at most quadratic in momentum}

For a concrete example, let us apply the above general formalism to a single particle subjected to external potentials so that the classical Hamiltonian takes the following form:
\begin{equation}
\underline{H}(q,\underline{p})=\frac{g^{ij}(q)}{2}(\underline{p}_i-A_i)(\underline{p}_j-A_j)+V, 
\label{classical Hamiltonian}
\end{equation}
where $A_i$, $i=x,y,z$ and $V(q)$ are vector and scalar potentials respectively, $g^{ij}(q)$ may depend on the configuration, and summation over repeated indices are assumed. Inserting  into the left equation of (\ref{Hamilton equation}) one gets 
\begin{equation}
\dot{q}^i(q,\underline{p};t)=\frac{\partial\underline{H}}{\partial\underline{p}_i}=g^{ij}(\underline{p}_j-A_j). 
\label{classical velocity field HPF particle in potentials}
\end{equation}
Substituting this into Eq. (\ref{classical continuity equation}), one has 
\begin{equation} 
\partial_t\Omega+\partial_{q_i}\Big((g^{ij}(\underline{p}_j-A_j))\Omega\Big)=0. 
\label{continuity equation particle in potentials}
\end{equation}
On the other hand, from Eq. (\ref{classical velocity field HPF particle in potentials}), $\theta(S)$ of Eq. (\ref{effective velocity divergence}) is thus given by
\begin{equation}
\theta(S)=\partial_{q_i}g^{ij}(\partial_{q_j}S-A_j). 
\end{equation} 
Using the above equation, Eq. (\ref{fundamental equation rederived}) becomes
\begin{eqnarray}
\underline{p}=\partial_qS+\frac{\lambda}{2}\frac{\partial_q\Omega}{\Omega},\hspace{20mm}\nonumber\\
-\underline{H}(q,\underline{p})=\partial_tS+\frac{\lambda}{2}\frac{\partial_t\Omega}{\Omega}+\frac{\lambda}{2}\partial_{q_i}g^{ij}(\partial_{q_j}S-A_j).
\label{fundamental equation particle in potentials} 
\end{eqnarray}

Now let us combine Eq. (\ref{continuity equation particle in potentials}) and pair of equations in (\ref{fundamental equation particle in potentials}). Substituting the first equation of Eq. (\ref{fundamental equation particle in potentials}) into Eq. (\ref{continuity equation particle in potentials}), one has 
\begin{equation}
\partial_t\Omega+\partial_{q_i}\Big(g^{ij}(\partial_{q_j}S-A_j)\Omega\Big)+\frac{\lambda}{2}\partial_{q_i}(g^{ij}\partial_{q_j}\Omega)=0. 
\label{FPE particle in potentials}
\end{equation}
On the other hand, inserting the upper equation of (\ref{fundamental equation particle in potentials}) into Eq. (\ref{classical Hamiltonian}) and plugging into the left hand side of the lower equation of (\ref{fundamental equation particle in potentials}), one has, after arrangement 
\begin{eqnarray}
\partial_tS+\frac{g^{ij}}{2}(\partial_{q_i}S-A_i)(\partial_{q_j}S-A_j)+V\hspace{30mm}\nonumber\\
-\frac{\lambda^2}{2}\Big(g^{ij}\frac{\partial_{q_i}\partial_{q_j}R}{R}+\partial_{q_i}g^{ij}\frac{\partial_{q_j}R}{R}\Big)\hspace{30mm}\nonumber\\
+\frac{\lambda}{2\Omega}\Big(\partial_t\Omega+\partial_{q_i}\Big(g^{ij}(\partial_{q_j}S-A_j)\Omega\Big)+\frac{\lambda}{2}\partial_{q_i}(g^{ij}\partial_{q_j}\Omega)\Big)=0,
\label{HJM particle in potentials 0}
\end{eqnarray}
where we have defined $R\doteq\sqrt{\Omega}$ and used the identity: $(1/4\Omega^2)\partial_{q_i}\Omega\partial_{q_j}\Omega=(1/2\Omega)\partial_{q_i}\partial_{q_j}\Omega-(1/R)\partial_{q_i}\partial_{q_j}R$. Inserting Eq. (\ref{FPE particle in potentials}), the last line vanishes to give
\begin{eqnarray}
\partial_tS+\frac{g^{ij}}{2}(\partial_{q_i}S-A_i)(\partial_{q_j}S-A_j)+V\nonumber\\
-\frac{\lambda^2}{2}\Big(g^{ij}\frac{\partial_{q_i}\partial_{q_j}R}{R}+\partial_{q_i}g^{ij}\frac{\partial_{q_j}R}{R}\Big)=0.
\label{HJM particle in potentials}
\end{eqnarray}
The above equation reduces into the Hamilton-Jacobi-Madelung equation when $g^{ij}$ is independent of $q$ and $\lambda=\pm\hbar$. We have thus the pair of coupled equations (\ref{FPE particle in potentials}) and (\ref{HJM particle in potentials}) which are parameterized by $\lambda$. 

Let us proceed to assume that $\Omega(q,\lambda;t)$ has the following symmetry: 
\begin{equation}
\Omega(q,\lambda;t)=\Omega(q,-\lambda;t),
\label{amplitude symmetry} 
\end{equation}
so that the probability density of $\lambda$ is unbiased $P(\lambda)=\int dq\Omega(q,\lambda;t)=P(-\lambda)$.
In this case, $S(q,\lambda;t)$ and $S(q,-\lambda;t)+S_0(\lambda)$, where $S_0(\lambda)$ is independent of $q$ and $t$, satisfy the same differential equation of (\ref{HJM particle in potentials}): namely the last term of Eq. (\ref{HJM particle in potentials}) is not sensitive to the signs of $\lambda$. Hence, if initially one has $S(q,\lambda;0)=S(q,-\lambda;0)+S_0(\lambda)$, then one will have 
\begin{equation}
S(q,\lambda;t)=S(q,-\lambda;t)+S_0(\lambda).
\label{phase symmetry} 
\end{equation}
$S_0(\lambda)$ is thus an odd function of $\lambda$, $S_0(-\lambda)=-S_0(\lambda)$ \cite{note}. The above properties can then be used to eliminate the last term of Eq. (\ref{FPE particle in potentials}): taking the case when $\lambda$ is positive add to it the case when $\lambda$ is negative and dividing by two, one gets 
\begin{equation}
\partial_t\Omega+\partial_{q_i}\Big(g^{ij}(\partial_{q_j}S-A_j)\Omega\Big)=0. 
\label{QCE particle in potentials}
\end{equation}
We have thus the pair of coupled equations (\ref{HJM particle in potentials}) and (\ref{QCE particle in potentials}) which are still parameterized by $\lambda$. 

Next, since $\lambda$ is non-vanishing, one can define the following complex-valued function:
\begin{equation}
\Psi(q,\lambda;t)\doteq R\exp\Big(\frac{i}{|\lambda|}S\Big). 
\label{general wave function}
\end{equation}
It differs from the Madelung transformation in which $S$ is divided by $|\lambda|$ instead of $\hbar$. The probability density for $q$ is thus given by 
\begin{equation}
\rho(q;t)=\int d\lambda\Omega=\int d\lambda|\Psi|^2. 
\label{generalized Schroedinger equation}
\end{equation}
Equations (\ref{HJM particle in potentials}) and (\ref{QCE particle in potentials}) can then be recast into the following modified Schr\"odinger equation: 
\begin{equation}
i|\lambda|\partial_t\Psi=\frac{1}{2}(-i|\lambda|\partial_{q_i}-A_i)g^{ij}(q)(-i|\lambda|\partial_{q_j}-A_j)\Psi+V\Psi. 
\label{generalized Schroedinger equation particle in potentials}
\end{equation}
Here we have assumed that the fluctuations of $\lambda$ in space and time are ignorable as compared to that of $S$. Let us then assume that $\Omega$ is factorizable as $\Omega(q,\lambda;t)=\rho(q,|\lambda|;t)P(\lambda)$, where $P(\lambda)$ takes the form given by Eq. (\ref{God's coin}). In this case, Eq. (\ref{generalized Schroedinger equation particle in potentials}) reduces into the celebrated Schr\"odinger equation
\begin{eqnarray}
i\hbar\partial_t\Psi_Q(q;t)=\hat{H}\Psi_Q(q;t),\hspace{20mm}\nonumber\\
\Psi_Q(q;t)\doteq\sqrt{\rho(q,\hbar;t)}e^{\frac{i}{\hbar}S_Q(q;t)}\hspace{2mm}
\&\hspace{2mm} S_Q(q;t)\doteq S(q,\pm\hbar;t), 
\label{Schroedinger equation particle in potentials} 
\end{eqnarray}
where the quantum Hamiltonian $\hat{H}$ is given by 
\begin{equation}
\hat{H}=\frac{1}{2}(\hat{p}_i-A_i)g^{ij}(q)(\hat{p}_j-A_j)+V,\hspace{2mm}\mbox{with}\hspace{2mm}\hat{p}_i\doteq -i\hbar\partial_{q_i}. 
\label{quantum Hamiltonian particle in potentials}
\end{equation}
From Eq. (\ref{Schroedinger equation particle in potentials}) we know that the Born's statistical interpretation of wave function is valid by construction for all time, $\rho(q;t)=|\Psi_Q(q;t)|^2$. In this context, Eq. (\ref{generalized Schroedinger equation}) should be regarded as a generalization of Born's rule. One can also see that unlike canonical quantization, the present statistical model of quantization selects a unique operator ordering in which $g^{ij}(q)$ is sandwiched by $\hat{p}-A$. 
 
The Schr\"odinger equation is thus a subset of the present model corresponding to a discrete unbiased random variable $\lambda=\pm\hbar$. It is then interesting to mention Ref. \cite{Gaveau analytic continuation} that the master equation of a particle moving with a fixed velocity, subjected to a random complete reverse of direction following a Poisson distribution, can be written into a Dirac equation (in the same way that the Schr\"odinger equation is connected to the dynamics of Brownian motion) through analytic continuation. Further, while $\lambda=\pm\hbar$ is a binary random variable, it might be a function of a set of continuous ``hidden'' variables $\xi=\{\xi_1,\xi_2,\xi_3,\dots\}$: $\lambda=f(\xi)$. For example, one may assume that $\lambda=\sqrt{\xi_1^2+\xi_2^2+\xi_3^2}=\pm\hbar$. Hence, $\xi=\{\xi_1,\xi_2,\xi_3\}$ lies on the surface of three-dimensional ball of radius $\hbar$. Let us divide the ball into two with equal area and attribute $\pm\hbar$ to each division. If $\xi$ moves on the surface sufficiently chaotically, then one will obtain $\lambda=\pm\hbar$ with equal probability. 
  
Since $\Omega$ and $S$ are multi-valued functions of $q$ due to their dependence on $\lambda$, then $\Psi(q,\lambda;t)$ is also a multivalued function of $q$. Let us consider the case when $\lambda$ can only take binary values $\lambda=\pm\hbar$ as assumed in Eq. (\ref{God's coin}). Equations (\ref{amplitude symmetry}) and (\ref{phase symmetry}) then become
\begin{equation}
\Omega(q,\hbar)=\Omega(q,-\hbar)\hspace{2mm}\&\hspace{2mm}S(q,\hbar)=S(q,-\hbar)+S_0(\hbar). 
\end{equation}
Hence, the probability density becomes single-valued while the phase is still multi-valued in accord with quantum theory. In contrast to quantum theory, the wave function is however multi-valued satisfying
\begin{equation}
\Psi_Q(q,\hbar)=\Psi_Q(q,-\hbar)e^{\frac{i}{\hbar}S_0(\hbar)}. 
\end{equation}
Assuming that the wave function is single-valued $\Psi_Q(q,\hbar)=\Psi_Q(q,-\hbar)$ one then has 
\begin{equation}
S_0=nh,\hspace{2mm}n=0,\pm 1,\pm 2,\dots,
\end{equation}
where $h=2\pi\hbar$ is the Planck constant. Hence, the set of single-valued quantum wave functions is a subset of the wave function of the present model when one initially picks a phase satisfying $S(q,\hbar;0)=S(q,-\hbar;0)+nh$, $n=0,\pm 1,\dots$. 

Wallstrom then argued in Ref. \cite{Wallstrom objection} that the Schr\"odinger equation with multi-valued wave functions does not give discrete quantum number, say discrete values of angular momentum. As discussed in Ref. \cite{AgungSMQ1}, in our model, a system can indeed possess any continuum values of (effective) angular momentum, energy etc. As shown there, however, a measurement of angular momentum can only give discrete possible values, that is one of the eigenvalue of quantum angular momentum operator as in standard quantum theory. Namely, quantum discreteness is a feature of measurement. 

Further, inserting the upper equation of (\ref{fundamental equation particle in potentials}) into Eq. (\ref{classical velocity field HPF particle in potentials}) one has
\begin{equation}
\dot{q}^i(\lambda)=g^{ij}(\partial_{q_j}S-A_j)+\frac{\lambda}{2}g^{ij}\frac{\partial_{q_j}\Omega}{\Omega}. 
\end{equation} 
One can then define an effective velocity as
\begin{equation}
\widetilde{v}^i(\lambda)\doteq\frac{\dot{q}(\lambda)+\dot{q}(-\lambda)}{2}=g^{ij}(\partial_{q_j}S-A_j), 
\label{effecive velocity field HPF particle in potentials}
\end{equation}
where the second equality is obtained by the virtue of Eqs. (\ref{amplitude symmetry}) and (\ref{phase symmetry}). Putting $\lambda=\pm\hbar$, it reduces into 
\begin{equation}
\widetilde{v}^i(\hbar)=g^{ij}(\partial_{q_j}S_Q-A_j)=\widetilde{v}^i(-\hbar). 
\label{effecive velocity field HPF particle in potentials}
\end{equation}
This is just the actual velocity of the particle in Bohmian mechanics, satisfying the continuity equation
\begin{equation}
\partial_t\rho+\partial_q\cdot(\rho\widetilde{v})=0. 
\end{equation}
Hence, we have an effective picture that the particle moves as if it is guided by the wave function \`a la Bohmian mechanics (de Broglie-Bohm pilot-wave theory). However, unlike the latter, the wave function in the present statistical model is not physically real. It is just an artificial mathematical tool to describe the dynamics and statistics of ensemble of copies of the system. Let us also mention that the effective velocity field $\widetilde{v}^i$ of Eq. (\ref{effecive velocity field HPF particle in potentials}) is equal to the ``naively observable velocity field'' reported in Ref. \cite{Wiseman naive velocity} obtained through weak measurement \cite{weak measurement} within the standard formalism of quantum mechanics.   

\section{Conclusion and discussion}

We have thus argued that quantization may be understood as the implication that the principle of stationary action is valid only approximately, and the distribution of the deviation from the stationary action along infinitesimal segment of path for a certain microscopical time scale $\tau_Q$ follows the exponential law given in Eq. (\ref{postulate of exponential distribution}). Moreover, in this statistical model, the absolute value of the random variable $|\lambda|=\hbar$ postulated in Ref. \cite{AgungSMQ1} is interpreted as the average deviation from the infinitesimal stationary action. The origin of Bell-non-locality in many particles system may then be traced back to the fact that $dS$ is evaluated along a segment of trajectory in configuration space rather than in ordinary space. Namely, a probability density is attributed to the infinitesimal action along a segment of trajectory in configuration space, rather than to the trajectory of each particle in ordinary space.   

While we have argued that the reduced Planck constant $\hbar$ is the average deviation from the infinitesimal stationary action in a certain microscopical time scale $\tau_Q$, the model does not offer physical mechanism that determines the value of the Planck constant. An attempt to provide a cosmic origin of the value of Planck constant is proposed by Calogero in Ref. \cite{Calogero conjecture}. Let us also mention the work by de la Pena and Cetto reported in Ref. \cite{de la Pena Planck C} which is partially motivated by the work of Calogero attempting to explain the value of Planck constant within the stochastic electrodynamics \cite{de la Pena SED}. 

Nor the present statistical model explains why the deviation from the infinitesimal stationary action follows an exponential law given by Eq. (\ref{postulate of exponential distribution}) rather than other distributions. To discuss this issue in the context of our model, it is then necessary to investigate the physics beyond the quantum mechanical time-scale $\tau_Q$. In particular, it is interesting to ask if one can devise a deterministic dynamical system which is valid in time scale less than $\tau_Q$, leading effectively in the quantum mechanical time scale of $\tau_Q$ to a stochastic behavior governed by Eq. (\ref{postulate of exponential distribution}) rather than other distributions. A proposal that quantum mechanics is emergent from a deterministic model for example is given in Ref. \cite{tHooft deterministic model}. To this end, it is also interesting to study the relation between the statistical model presented in this paper and that developed in Ref. \cite{AgungSMQ0} based on Hamilton-Jacobi theory with a specific random constraint determined uniquely by the Lagrangian. 

The present statistical model suggests several possible generalizations of the Schr\"odinger equation of Eq. (\ref{Schroedinger equation particle in potentials}). The first immediate possible generalization is given by Eq. (\ref{generalized Schroedinger equation particle in potentials}) which is valid even when the distribution of $\lambda$ deviates from Eq. (\ref{God's coin}) but is still unbiased $P(\lambda)=P(-\lambda)$. For example, one assumes that $\lambda$ fluctuates around $\pm\hbar$ with a very small width. Such type of generalization is discussed in Ref. \cite{AgungSMQ2} and is shown to lead to testable possible corrections to the statistical prediction of quantum mechanics. The other type of generalization is to choose different form of $\theta(S)$ which still reduces into Eq. (\ref{classical velocity field}) in the limit $S\rightarrow\underline{S}$ to guarantee smooth classical correspondence, and gives Eq. (\ref{effective velocity divergence}) as its first order approximation. Yet another possibility is to take different distribution of infinitesimal action which reduces into Eq. (\ref{principle of stationary action}) in some physically reasonable limit. All the above mentioned problems together with the application of the present statistical model to foundational problems of quantum mechanics will be the object of future investigation. 
 
\begin{acknowledgments} 

This research was initiated when the author held an appointment with RIKEN under the FPR program.  

\end{acknowledgments}

\end{document}